\documentclass[notitlepage,twocolumn,twoside]{article}
\usepackage{graphicx}
\usepackage{url}
\begin{document}
\title{Return on Investment Driven Observability}
\author{
Michael~Hausenblas~\texttt{mh9@o11y.engineering}
\thanks{This is an article published in March 2023 and distributed under Creative Commons
Attribution 4.0 International (CC BY 4.0) license and is intended for observability practitioners. 
The opinions expressed inhere are entirely the ones of the author and by no means imply that 
his employer, Amazon Web Services, endorses or supports them.}
}
\maketitle
\begin{abstract}
Observability, in cloud native systems, is the capability to continuously 
generate and discover actionable insights, based on signals from the system under observation. 
How do you know what insights are the most useful ones? What signals should you be using 
to generate insights? This article discusses challenges arising when rolling out observability
in organizations and how you can, based on Return on Investment (RoI) analysis, address said challenges. 
\end{abstract}

\section{Introduction}
\begin{figure}[!htbp]
\centering
\includegraphics[width=2.5in]{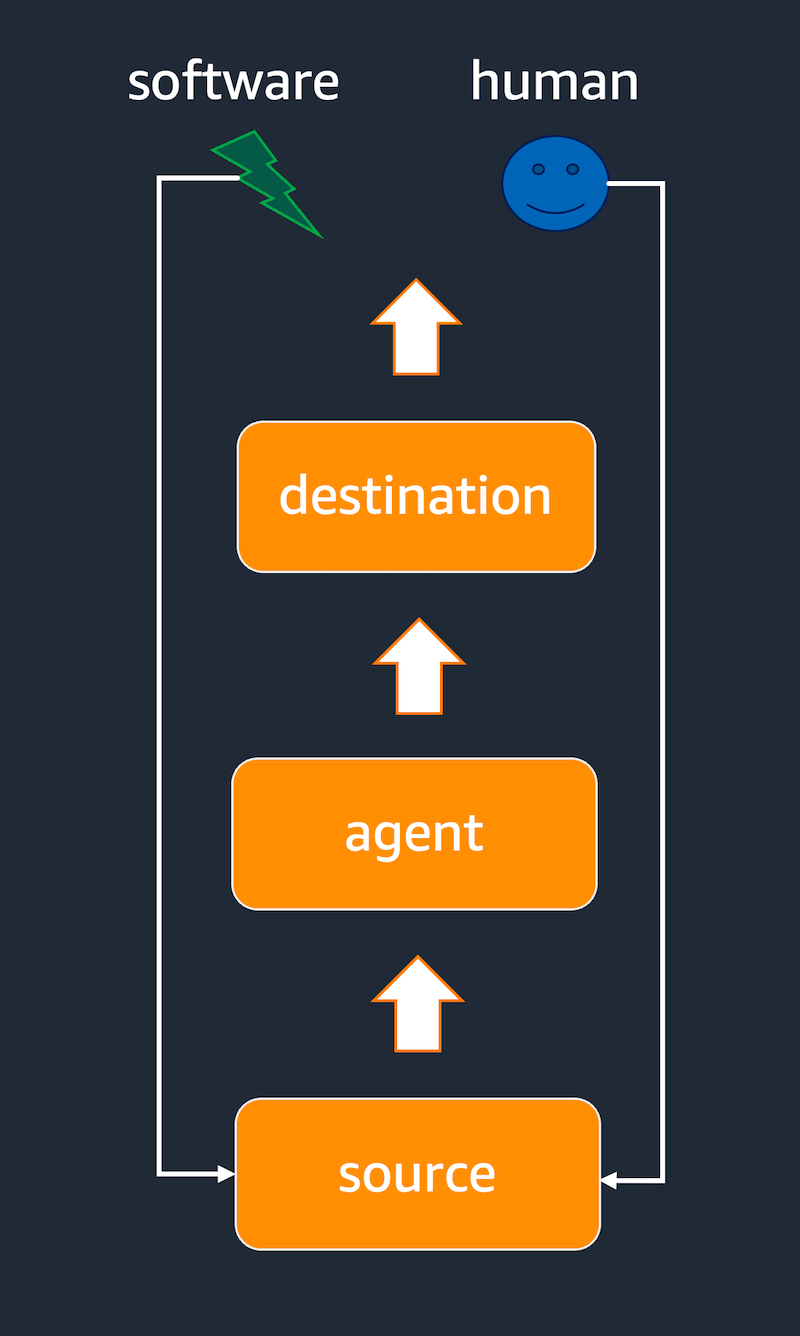}
\caption{The observability feedback loop.}
\label{figO11yFeedbackLoop}
\end{figure}

In cloud native systems~\cite{CNCF:o11y}, observability is the capability to continuously 
generate and discover actionable insights based on signals from the system under observation (SOT). 
The Fig.~\ref{figO11yFeedbackLoop} shows the overall feedback loop with the sources at the bottom
representing the SOT and the human or software at the top as the entities consuming signals.

While on a high level this is fine, practitioners often find it hard to apply the idea or even
introduce observability in their organization. Too many things, such as the business value
or the necessary effort to gather the insights remain unclear or at best we end up with
vague guidance, not uncommon from vendors who prioritize positioning their offerings over
talking about the hard bits.

With cloud native systems we mean either containerized (micro-)services orchestrated by Kubernetes,
Nomad, or proprietary systems such as Amazon Elastic Container Service (ECS) and AWS Lambda. Note
that we don't require the services to run in a public cloud setting, in other words, if you
run a Kubernetes cluster on-premises you may equally interested in and are affected by this article's
considerations. The main properties of cloud native systems are small units of deployments (pods, tasks)
and the short-lived nature of said units, compared to a monolith. These properties cause challenges,
including discovery of sources, temporal gaps, volume, etc.

So let's address the elephant in the room: to arrive at actionable insights, you first have to invest.
Investments come in different shapes, as come the returns, so that's what we will  have a look
at in section~\ref{secRoI}. Then, in section~\ref{secWB} we will dive into how to identify and
establish meaningful insights, based on goals, be that technical ones or business goals. 
In section~\ref{secSelSi} we then discuss how to go about selecting signals to drive the insights,
based on their costs.

The article is a companion of and my book "Cloud Native Observability in Action"~\cite{MH9:CNO11YIA}
and the idea is to provide a condensed treatment of the RoI-driven aspect of how to approach and implement 
observability. We will specifically describe how to use OpenTelemetry~\cite{CNCF:OTel} to address
the challenges presented.

\label{secRoI}
\section{What is RoI?}
Whenever there is a thing of value, to the organization or to an individual, the question that 
inevitably comes along is: What are the costs associated with it? Let me give you an example: say,
you're considering to install solar panels. That's a good thing, right? Now, you will have to invest
time and money to make this happen, fair enough. What do you get in return? A number of things, including
lower electricity bills, more independence of the grid, a contribution to fight climate disaster,
just to name a few benefits. What you've just seen is Return of Investment (RoI)~\cite{IP:RoI} in action.

With observability it's just like that: you need to invest to gain something. The RoI can take many shapes:
from cost savings to being able to ship faster to selling more stuff. What is important to remember is
that there is no such thing as a free lunch.

The investments you will be facing can revolve around:

\begin{itemize}
  \item \textbf{Sources}: Instrumentation of the code, requiring manual work or the overhead of auto-instrumentation,
  impacting latency and resource usage.
  \item \textbf{Agents}: Running, patching, upgrading, scaling, and configuring agents that collect, process,
  and ingest signals into observability back-ends.
  \item \textbf{Destinations}: In general the costs break down into ingest, storage, and query. Different providers
  offer different pricing models from charge per event/host/service to per seat to ingest or query based.
\end{itemize}

You will also come across buy vs. build and the considerations to use open source (components) either
directly in your solution or via a managed open source offering from your cloud provider of choice. As
discussed in~\cite{MH9:GHRM}, open source is a great way to collaborate on software and comes with many advantages from security (many eyes
principle) to being able to troubleshoot more effectively, because you have the source code available
to you. There are also challenges with using open source, specifically the necessary commitment to
keep up with the (release) pace of the open source project(s) you plan to use in your solution. 

Investments covers everything you need to commit to do or willing to spend in terms of time or money, but 
what are returns? Returns can include but are not limited to:

\begin{itemize}
  \item By empowering developers to test and validate code in the context of their IDE, they are able to 
  iterate faster, shipping new features or bug fixes more quickly. This developer observability is an 
  emerging field, making continuous profiling, distributed tracing, and metrics on and for the source code
  readily available to developers at the time they implement new features or provide a bug fix.
  \item Using fewer resources for the same functionality. This can be relevant for both developers and
  operations roles, typically based on profiles (CPU, memory usage and execution times).
  \item In case of an incident, you can troubleshoot faster and recover from a failure more quickly.
  \item By being transparent, keeping internal and external customers in the loop
  on what is going on, observability can result in happier and more loyal customers.
\end{itemize}

Now that we have a basic understanding of the necessary investments and potential returns, let us
shift our focus on how to establish goals and how to go from there to actionable insights.

\label{secWB}
\section{From Goals To Insights}
Let's first define what actionable insights mean and then discuss how to work backwards from goals. The basic
idea is to identify what brings the most value to the organization, helping to align stakeholders from
business (sales, product, marketing, business development, etc.) and IT (ops, dev, security, networking, etc.)
roles. This could be in green field environments (for example moving to the cloud) or in a brown field
setup which often comes with requirements to integrate with existing monitoring or compliance tooling.

\subsection{What Does Actionable Mean?}
An actionable insight literally means that. You can action on it. For example, if you come across an
error message like:
\begin{verbatim}
ERROR 142. Operation did not succeed!     
\end{verbatim}
then that's the opposite of actionable. It neither tells you why the error happened nor how you can fix it. 
In contrast, an actionable insight would look something like this:
\begin{verbatim}
Disk is full and can't write file, 
please make some space available.
\end{verbatim}
Now, \textbf{who} can action on the insights generated from telemetry signals? In general the consumption
of the insights can be through:
\begin{itemize}
  \item humans, such as an on-call engineer being paged and looking at a dashboard, and
  \item machines, be that traditional software (think: cluster auto-scaler) or applying
  machine learning to  telemetry data, for example to perform anomaly detection~\cite{GOOG:ML}.
\end{itemize}
We now know what actionable insights mean and also for whom and move on to identifying useful insights.

\subsection{Identify Actionable Insights}
 How do you know what insights are the most useful ones? TL;DR: agree on the goal is and work backwards 
from it to understand what insights you need. Goals could be either business-related or
technical goals, including but not limited to:
\begin{enumerate}
  \item In case of an incident, reducing the Mean Time To Restore (MTTR)~\cite{SPLUNK:MTTR,NR:MTTR,ES:MTTR}.
  \item For developers, increasing the velocity (for example, shipping more features in a sprint).
  \item Enable resource usage optimization flows (based on profiles or traces).
  \item Business-related: increase conversion rate and with it sales, reduce check-out time, increase the shopping basket item count.
\end{enumerate}
No matter the type and scope of the goal you're after: start there. In other words, if you don't have
a clear goal in mind, observability can quickly become a end in itself and you risk
being perceived as a cost center rather than a profit center. 

OK, so we have an idea about what insights we're after to achieve a goal. But what signals
are the right ones to use to drive the insights? Let's see!

\label{secSelSi}
\section{Selecting Signals}
How to go about selecting signals? In a nutshell: based on their costs/risks and value.

\subsection{The Costs of Signals}
Before we get to the costs, risks, and value of the four major signal
types I will point out that you should consider standardizing on the
telemetry (instrumentation, format, metadata for correlation) and the good news
is that in 2023 there is a de-facto industry standard available to you: 
OpenTelemetry (OTel)~\cite{CNCF:OTel}.

\label{secCostOfLogs}
\subsubsection{Logs}
In Table~\ref{tabCostsLogs} on page \pageref{tabCostsLogs} you see the costs of logs along with
mitigation strategies to address them.

\begin{table}
\caption{The costs and risks of logs}
\label{tabCostsLogs}
\centering
\begin{tabular}{|r||l|l|}
\hline
& \emph{risk} & \emph{mitigation} \\
\hline
\textbf{instrumentation} &  low fidelity & traces \\
\hline
\textbf{agents} & standards & OTel \\
\hline
\textbf{destinations} & volume & data temp. \\
\hline
\end{tabular}
\end{table}
Logs are used everywhere and focus on a particular service. The fidelity is in general low since
developers typically emit log lines ad-hoc, providing output at unknown locations of the execution
flow within a service. There are many standards in the logs domain (Syslog, SIEM, etc.) and at the same time
there is no one dominating standard. Users often report that the sheer volume aof logs is
challenging and with it the access times are usually the focus (cf. Loki and Zink).

To manage the volume in the destination, you want to differentiate between online (or: hot) data
that has to be accessible with minimal latency, less than 100 ms and offline (or: cold) data
that you want to keep around for long-term analysis but you are fine if it takes
seconds or minutes to access, for example through online services such as Amazon S3 Glacier.
This technique is called data temperature management.

\label{secCostOfMetrics}
\subsubsection{Metrics}
In Table~\ref{tabCostsMetrics} on page \pageref{tabCostsMetrics} you see the costs of metrics along with
mitigation strategies to address them.

\begin{table}
\caption{The costs and risks of metrics}
\label{tabCostsMetrics}
\centering
\begin{tabular}{|r||l|l|}
\hline
& \emph{risk} & \emph{mitigation} \\
\hline
\textbf{instrumentation} & which metrics &  usage analysis  \\
\hline
\textbf{agents} & pull vs push & OTel \\
\hline
\textbf{destinations} & cardinality & federation/LTS  \\
\hline
\end{tabular}
\end{table}
Metrics are often used and the Prometheus exposition format (OpenMetrics) has established itself 
as the cloud native standard to represent metrics on the wire (with OpenTelemetry being compatible and
inter-operable to OpenMetrics). With Prometheus, the pull-based approach (agent scrapes targets, in
Prometheus terminology) rather than applications pushing metrics to the agent, has become
popular. This makes a lot of sense in the context of cloud native systems with volatile workloads. 

It is typically challenging to decide what metrics to collect and
cloud native systems with many moving parts can be overwhelming, offering a torrent of
low level (and often low value) metrics readily available, from container-level to network
to storage.

With metrics, cardinality explosion~\cite{MH9:CNO11YIA} is a problem and so can aggregation of metrics be.
You want to work backwards from the usage of the metrics to selecting which ones you
want to collect (and ingest into back-ends). The rule of thumb is: if a metrics is neither
used in a dashboard nor alerted on, you have to have a very good reason to take it into
consideration. If you have a fleet of machines or (Kubernetes) clusters and want to consume metrics
in a central place and/or have long-term storage requirements, you want to consider horizontally
scalable Prometheus-compatible back-end such us CNCF Cortex or CNCF Thanos.

\label{secCostOfTraces}
\subsubsection{Traces}
In Table~\ref{tabCostsTraces} on page \pageref{tabCostsTraces} you see the costs of traces along with
mitigation strategies to address them.

\begin{table}
\caption{The costs and risks of distributed traces}
\label{tabCostsTraces}
\centering
\begin{tabular}{|r||l|l|}
\hline
& \emph{risk} & \emph{mitigation} \\
\hline
\textbf{instrumentation} & async services & SpanLinks  \\
\hline
\textbf{agents} & throughput & OTel\\
\hline
\textbf{destinations} & volume & sampling  \\
\hline
\end{tabular}
\end{table}
Distributed tracing enjoy some adoption (somewhere between 30\% and 40\% uptake overall) and for synchronous services
such as request/response services this is a solved challenge. With asynchronous services such as message busses,
one can find it challenges to gain a complete picture since context propagation is tricky and the
default semantics between root span and child spans is not applicable or even makes sense~\cite{MT:ASYNC}. OpenTelemetry
provides the \textit{SpanLink} feature to address these issues.

You can employ sampling to control the volume of your spans using different techniques such as head-based
and tail-based sampling~\cite{PAA:DTIP}. In order to reduce costs in the write path, in-memory caching of recently
used spans is vital. Also, think about how you will index and look up traces. The way users can search for or 
filter (along services, error codes, etc.) can make the difference between rapid adoption and an expensive experiment.

\label{secCostOfProfiles}
\subsubsection{Profiles}
In Table~\ref{tabCostsProfiles} on page \pageref{tabCostsProfiles} you see the costs of profiles along with
mitigation strategies to address them.

\begin{table}
\caption{The costs and risks of profiles}
\label{tabCostsProfiles}
\centering
\begin{tabular}{|r||l|l|}
\hline
& \emph{risk} & \emph{mitigation} \\
\hline
\textbf{instrumentation} & collection & automation \\
\hline
\textbf{agents} & overhead & kernel \\
\hline
\textbf{destinations} & indexing & open standards\\
\hline
\end{tabular}
\end{table}

Continuous profiling is a relative new domain and hence, compared to the previous three signal
types, fewer good practices exist. What seem clear is that you want to automate the collection
(using eBPF and auto-instrumentation) rather than manually emitting profiles from your code.

In dev/test continuous profiling can be used without any worries, for production environments
you need to take the overhead of the profiler and the necessary debug symbols into account.
In any case, if you plan to invest in this area you should strongly consider to build on
open standards such as \texttt{pprof} and OpenTelemetry. For profiling back-ends there is
an array of research and open source implementations (such as from Polar Signals~\cite{PS:PDB})
available.

\section{Conclusion}

We covered what actionable insights are and how to work backwards from business or technical goals.
Things you should consider when starting off on your observability journey:

\begin{itemize}
  \item Agree with your stakeholders what goals you want to achieve with observability in your organization.
  \item Consider the costs involved and try to assess what the return is, communicating it clearly
  to your stakeholders.
  \item Buy versus build: you need to understand if building it yourself is a competitive 
  advantage for your organization or not.
  \item Open source and open standards are valuable and can help mitigating vendor lock-in, 
  however, you need to be willing to commit resources to track upstream.
  \item In cloud native systems, distributed traces are more valuable and useful than logs alone.
  \item Whenever possible, try to benefit from automation, for example, in the context of
  instrumentation but also correlation.
\end{itemize}

Thank you for reading and I'd be interested to hear your feedback and in learning from
you what experiences you made introducing observability in your organization, what
lessons learned you had, and what you would do differently now.


\newpage
\begin{thebibliography}{1}

\bibitem{CNCF:o11y}
Cloud Native Computing Foundation Glossary, \emph{Definition of the term 'observability'}, 
\url{https://glossary.cncf.io/observability/}, Last visited: 2023-03-16.

\bibitem{CNCF:OTel}
Cloud Native Computing Foundation,\emph{OpenTelemetry},
\url{https://opentelemetry.io/docs/}, Last visited: 2023-03-18.

\bibitem{IP:RoI}
Investopedia (part of the Dotdash Meredith publishing family), \emph{Return on Investment (ROI): How to Calculate It and What It Means}, 
\url{https://www.investopedia.com/terms/r/returnoninvestment.asp}, Last visited: 2023-03-17.

\bibitem{MH9:CNO11YIA}
Michael Hausenblas, \emph{Cloud Native Observability in Action}, 2023, ISBN 9781633439597, Manning.

\bibitem{MH9:GHRM}
Michael Hausenblas, \emph{Look beyond lock-in with open source observability}, 2022,
\url{https://github.com/readme/guides/open-source-observability/}, Last visited: 2023-03-23.

\bibitem{GOOG:ML}
Chun-Liang Li and Kihyuk Sohn, \emph{Discovering Anomalous Data with Self-Supervised Learning}, 2021, 
\url{https://ai.googleblog.com/2021/09/discovering-anomalous-data-with-self.html}, Last visited: 2023-03-23.

\bibitem{SPLUNK:MTTR}
Splunk, \emph{How to Use Observability to Reduce MTTR}, 2021, 
\url{https://www.splunk.com/en_us/blog/devops/using-observability-to-reduce-mttr.html}, Last visited: 2023-03-23.

\bibitem{NR:MTTR}
New Relic, \emph{Reducing MTTR the Right Way},  
\url{https://newrelic.com/devops/how-to-reduce-mttr}, Last visited: 2023-03-23.

\bibitem{ES:MTTR}
Elastic, \emph{Elastic Observability: Driving mean time to resolution to zero}, 2021, 
\url{https://www.elastic.co/blog/elasticon-global-observability}, Last visited: 2023-03-23.

\bibitem{MT:ASYNC}
Martin Thwaites, \emph{Understanding Distributed Tracing with a Message Bus}, 2023, 
\url{https://www.honeycomb.io/blog/understanding-distributed-tracing-message-bus}, Last visited: 2023-03-23.

\bibitem{PAA:DTIP}
Austin Parker, Daniel Spoonhower, Jonathan Mace, Ben Sigelman, and Rebecca Isaacs,
\emph{Distributed Tracing in Practice}, 2020, ISBN 9781492056621, O'Reilly.

\bibitem{PS:PDB}
Alfonso Subiotto Marques, \emph{Designing Your Indexes for Better Database Performance}, 2023, 
\url{https://www.polarsignals.com/blog/posts/2023/03/21/designing-your-indexes/}, Last visited: 2023-03-23.

\end{thebibliography}
\end{document}